# Multiple Weyl fermions in the noncentrosymmetric semimetal LaAlSi

Hao Su[1,2,3], Xianbiao Shi[4,5], Jian Yuan[1], Yimin Wan[6], Erjian Cheng[6], Chuanying Xi[7], Li Pi[7], Xia Wang[1,8], Zhiqiang Zou[1,8], Na Yu[1,8], Weiwei Zhao[4,5*], Shiyan Li[6,9*], Yanfeng Guo[1*]

[1] School of Physical Science and Technology, ShanghaiTech University, Shanghai 201210, China
[2] Shanghai Institute of Optics and Fine Mechanics, Chinese Academy of Sciences, Shanghai 201800, China
[3] University of Chinese Academy of Sciences, Beijing 100049, China
[4] State Key Laboratory of Advanced Welding & Joining and Flexible Printed Electronics Technology Center, Harbin Institute of Technology, Shenzhen 518055, China
[5] Key Laboratory of Micro-systems and Micro-structures Manufacturing of Ministry of Education, Harbin Institute of Technology, Harbin 150001, China
[6] State Key Laboratory of Surface Physics, Department of Physics, and Laboratory of Advanced Materials, Fudan University, Shanghai 200438, China
[7] Anhui Province Key Laboratory of Condensed Matter Physics at Extreme Conditions, High Magnetic Field Laboratory of the Chinese Academy of Sciences, Hefei 230031, China
[8] Analytical Instrumentation Center, School of Physical Science and Technology, ShanghaiTech University, Shanghai 201210, China
[9] Collaborative Innovation Center of Advanced Microstructures, Nanjing 210093, China

The noncentrosymmetric $R$Al$Pn$ ($R$ = rare earth, $Pn$ = Si, Ge) family, predicted to host nonmagnetic and magnetic Weyl states, provide an excellent platform for investigating the relation between magnetism and Weyl physics. By using high field magnetotransport measurements and first principles calculations, we have unveiled herein both type-I and type-II Weyl states in the nonmagnetic LaAlSi. By a careful comparison between experimental results and theoretical calculations, nontrivial Berry phases associated with the Shubnikov-de Haas oscillations are ascribed to the electron Fermi pockets related to both types of Weyl points located ~ 0.1 eV above and exactly on the Fermi level, respectively. Under high magnetic field, signatures of Zeeman splitting are also observed. These results indicate that, in addition to the importance for exploring intriguing physics of multiple Weyl fermions, LaAlSi as a comparison with magnetic Weyl semimetals in the $R$Al$Pn$ family would also yield valuable insights into the relation between magnetism and Weyl physics.

*Corresponding authors:
wzhao@hit.edu.cn,
shiyan_li@fudan.edu.cn,
guoyf@shanghaitech.edu.cn.

## I. INTRODUCTION

Realization of Weyl fermions in solids has aroused intense interest in condensed matter physics due to their fascinating physical properties associated with the exotic nontrivial topological state and profound potential applications in topological spintronics [1-3]. In solids, the Weyl fermions emerge in Weyl semimetals (WSMs) in the form of low energy excitations that can be viewed as massless fermionic quasiparticles and described by the Weyl equation with $2 \times 2$ complex Pauli matrices [4-6]. Specifically, the low energy excitations in the electronic band structure is in the vicinity of doubly degenerate band crossing point, namely, the Weyl point (WP), in the momentum space. The WPs carrying Abelian Berry curvature must exist in pairs of opposite chirality, which locate at isolated positions and are protected by a Chern number $\pm 1$ defined on a sphere enclosing the WP. Depending on the Femi surface (FS) geometry at the WPs, WSMs can be basically classified into type-I and type-II [7]. Type-I WSMs have a point-like FS at the vertical Weyl cone, while the type-II WSMs have tilted Weyl cone as a connector of the electron and hole pockets, which is capable of breaking the Lorentz invariance [8-12]. Due to the requirement of protection by different symmetries, the realization of different Weyl fermions in a single material is very difficult, which has therefore been scarcely reported. Considering the intriguing properties associated with different types of WSMs, the coexistence of type-I and type-II Weyl fermions in a single phase would promote unique quantum effects, making such WSMs extremely desirable.

On the other side, to establish the Weyl state, it is necessary to break either space-inversion (SI) or time-reversal (TR) symmetry. Up to date, most WSMs are realized through breaking the SI [4-6], whereas rather few have been verified to be realized through breaking the TR. For SI breaking WSMs, rich exotic physical phenomena have been observed, including the high mobility charge carriers [13], extremely large magnetoresistance [14,15], chiral anomaly negative magnetoresistance [16-18], anomalous nonlinear quantum Hall effect [19-21], and nonlocal gyrotropic effects [22], etc. Compared with SI breaking WSMs, TR breaking ones would find more opportunities for use in spintronics [23-26]. Taking the Kagome lattice $Fe_3Sn_2$ as an example, the distribution of WPs is determined by the spin texture, which could be controlled by the magnetization switching technique [27]. The very recently experimentally verified magnetic topological nodal-line semimetal $Co_2MnGa$ [28] and the magnetic WSM $Co_3Sn_2S_2$ [23,29] both are centrosymmetric ferromagnets. The nocentrosymmetric *RAlPn* (*R* = rare earth element, *Pn* = Si or Ge) which naturally lose the SI have been proposed as a new family of WSMs [30-38], covering nonmagnetic and magnetic Weyl fermions depending on the rare earth elements. The angle-resolved photoemission spectroscopy (ARPES) measurements unveiled both type I and type II Weyl states in nonmagnetic LaAlGe [30], providing a rare example waiting for further study on its possible intriguing properties associated with the unusual topological states. In the ferromagnetic PrAlGe, the ARPES measurements found surface topological arc corresponding to the projected charge of ±1 in the surface Brillouin zone [31]. Moreover, large anomalous Hall effect was reported in PrAlGe and CeAlGe, likely arising from the diverging bulk Berry curvature fields associated with the magnetic Weyl band structure [31,34,35]. In NdAlSi, the neutron diffraction and magnetotransport measurements, assisted by the theoritical calculations, found that the periodicity of long-wavelength magnetic order is linked to the nesting vector between two topologically nontrivial Fermi pockets, suggesting Weyl fermion driven collective magnetism [36]. The magnetic members present rare examples with breaking both the SI and TR symmetries, thus offering opportunities to study the interplay between magnetism and Weyl fermions. However, the nonmagnetic members in this family remain much less studied. To achieve more crucial insights into this issue, the study on the nonmagnetic members of *RAlPn* would serve as a sharp comparison with the magnetic WSMs in this family for providing useful clues.

In this paper, we present our high field magnetotransport measurements and first principles calculations of LaAlSi, a nonmagnetic member belonging to the *RAlPn* family. The high field magnetotransport measurements reveal strong quantum oscillations associated with nontrivial Berry phase. According to the calculations, there are three kinds of nonequivalent WPs, marked as $W_1$, $W_2$, and $W_3$, where $W_2$ belongs to type-II Weyl fermion locating exactly on the Fermi level $E_F$ and the $W_1$ and $W_3$ belong to type-I Weyl fermion distributing about 0.1 eV above $E_F$. The perfect agreement between the first principles calculations and the experimental measurements on angular dependence of FS cross sectional areas suggests that the electron pockets connected to the Weyl cones of both $W_1$ and $W_2$ WPs dominate the transport properties of LaAlSi.

## II. EXPERIMENTAL

The LaAlSi crystals were grown by using the self-flux method. Starting materials of La (99.95%, aladdin), Al (99.999%, aladdin) and Si (99.9999%, aladdin) blocks were mixed in a molar ratio of 1: 10: 1 and placed into an alumina crucible which was then sealed into a quartz tube in vacuum. The assembly was heated in a furnace up to 1100 °C within 10 hrs, kept at the temperature for 20 hrs, and then slowly cooled down to 750 °C at a temperature decreasing rate of 0.5 °C/h. The excess Al was removed at this temperature by quickly placing the assembly into a high-speed centrifuge. Black crystals with shining surface in a typical dimension of $1.7 \times 1.1 \times 0.3$ mm$^3$ were finally left in the alumina crucible, shown by the picture as an inset of Fig. 1(a).

The phase and quality examinations of LaAlSi were performed on the Bruker AXS D8 Advance powder crystal X-ray diffractometer with Cu K$_{\alpha 1}$ ($\lambda$ = 1.54178 Å) at 298 K. The Bragg reflections shown in Fig. 1(a) by the main panel can be nicely refined by using the space group $I4_1md$ (No. 109) as the initial model, consistent with those analogues reported earlier [34,37,38]. To check the result analyzed based on powder X-ray diffraction measurement, crystals from the same batch were measured on the Bruker D8 single crystal X-ray diffractometer with Mo K$_{\alpha 1}$ ($\lambda$ = 0.71073Å). The measurements were measured from room temperature to 150 K, which did not detect any structure phase transition. At 150 K, the best solutions of single crystal X-ray diffraction refinement supports the noncentrosymmetric space group of $I4_1md$ with lattice parameters $a = b = 4.3055(3)$ Å, $c = 14.664(1)$ Å, $\gamma = \alpha = \beta = 90°$. The consistence between powder and single crystal X-ray diffraction measurements guarantees the correct phase and high-quality of the crystals used in this study. Based on the structure refinement results, the schematic structure was drawn and shown by the right inset to Fig. 1(a). The structure is body-centered and consists of stacked La, Al and Si layers along the $c$ axis. The magnetotransport measurements, including the resistivity and Hall effect measurements, were carried out using a standard Hall bar geometry in a commercial DynaCool physical properties measurement system (PPMS) from Quantum Design. The magnetotransport was also measured in the steady high magnetic field facility at Hefei (SHMFF). We used two crystals from the same batch to repeat the measurements and got rather close results, thus guaranteeing the reliability of the measurements.

The first principles calculations were carried out within the framework of the projector augmented wave (PAW) method [39,40], and employed the generalized gradient approximation (GGA) [41] with Perdew-Burke-Ernzerhof (PBE) [42] formula, as implemented in the Vienna *ab initio* Simulation Package (VASP) [43-45]. For all calculations, the cutoff energy for the plane-wave basis was set as 500 eV, the Brillouin zone sampling was done with a Γ-centered Monkhorst-Pack k-point mesh of size 8 × 8 × 8, and the total energy difference criterion was defined as $10^{-8}$ eV for self-consistent convergence. The optimized structural parameters were used in the electronic structure calculations. The maximally localized Wannier functions (MLWFs) [46-48] for La $d, f$; Al $s, p$ and Si $s, p$ orbitals were constructed to determine the hopping parameter values for tight-binding model. The Wannier Tools package [49], which works in the tight-binding framework was used to compute the surface spectrum including the surface density of states and the Fermi arcs based on the iterative Green's function method [50].

## III. RESULTS AND DISCUSSION

The temperature ($T$) dependence of longitudinal resistivity $\rho_{xx}$ measured with electrical current $I$ along the (001) plane at a magnetic field $B = 0$ T is presented by the inset to Fig. 1(c), displaying a typical semi-metallic conduction. The MR, defined as MR = [$\rho(B) - \rho(0)$]/$\rho(0)$ × 100% in which $\rho(B)$ and $\rho(0)$ represent the resistivity with and without $B$, respectively. Fig. 1(b) shows the $B$ dependence of the longitudinal resistivity $\rho_{xx}$ with the magnetic field $B$ along the [001] direction in the magnetic field range of -9 T – 9 T. The MR vs. $B$ in the temperature range of 2 - 150 K shows a quadratic-like change at very low magnetic field while nearly linear evolution with increasing $B$, and the MR reaches ~ 66% at 9 T. The Hall resistivity, seen in Fig. 1(c), exhibits a nearly linear dependence on $B$ with a negative slope, indicating that dominant role of electron carriers for the transport. The result agrees well with the theoretical calculations presented later. By a linear fit to the Hall resistivity with the single band model, i. e. $\rho_{xy} = R_H \cdot B$, where $R_H = 1/en$ is the Hall coefficient, $n$ donates the carrier density and $e$ is the electron charge, the obtained temperature dependence of carrier density $n$ shown in Fig. 1(d) seems almost temperature independent with a rather slight change over the measured temperature range. Then the temperature dependent carrier mobility $\mu_e$ is obtained by using $\sigma = ne\mu_e$, which shows nearly a monotonic decrease with increasing the temperature.

Unfortunately, the low-field MR only displays very weak quantum oscillations that are rather difficult for further analysis, so the magnetotransport measurements were then subjected to high magnetic field up to 35 T. Shown by the main panel of Fig. 2(a), striking Shubnikov-de Haas (SdH) quantum oscillations in the MR are visible, thus allowing further analysis. After carefully subtracted a fourth-power polynomial background, the SdH oscillations at different temperatures from 2 to 50 K against the reciprocal magnetic field 1/$B$ are presented in Fig. 2(b), which could be well described by the Lifshitz-Kosevich (L-K) equation [51]:

$$\Delta\rho \sim R_S R_T R_D \cos\left[2\pi\left(\frac{F}{B} + \gamma - \delta\right)\right],$$

Where $R_S = \cos(\pi g m^*/2m_e)$, $R_T = 2\pi^2 k_B T/\hbar\omega_c / \sinh(2\pi^2 k_B T/\hbar\omega_c)$, and $R_D = \exp(-2\pi^2 k_B T_D/\hbar\omega_c)$ represent the damping factors due to spin splitting, temperature and scattering, respectively. In the expressions of $R_T$, $R_T$ and $R_D$, $m^*$ is the effective cyclotron mass, $m_e$ denotes the free electron mass, $k_B$ is the Boltzmann constant, $\hbar$ is the Planck's constant, $F$ is the oscillation's frequency, $\gamma - \delta$ is the phase shift, $\omega_c = eB/m^*$ is the cyclotron frequency, $T_D$ is the Dingle temperature defined by $T_D = \hbar/2\pi k_B \tau_Q$ with $\tau_Q$ being the quantum scattering lifetime. The fast Fourier transform (*FFT*) spectra of the SdH oscillations, depicted in Fig. 2(c), disclose two fundamental frequencies at $F_\alpha$ = 7.96 T and $F_\beta$ = 47.78 T. The corresponding external cross-sectional areas of the FSs are $A$ = 0.0758 and 0.455 nm$^{-2}$, calculated by using the Onsager relation $F = (\hbar/2\pi e)A$. The small effective cyclotron mass $m^*$ at the $E_F$ could be obtained by fitting the temperature dependence of the *FFT* magnitude by the temperature damping factor $R_T$, as is shown in Fig. 2(d), giving $m_\alpha^*$ = 0.045 $m_e$ and $m_\beta^*$ = 0.049 $m_e$. The Fermi wave vectors are estimated to be 0.1554 and 0.381 nm$^{-1}$ from $k_F = \sqrt{2eF/\hbar}$ and the very large Fermi velocities $v_F$ = 4×10$^5$ and 9×10$^5$ m s$^{-1}$ are calculated from $v_F = \hbar k_F/m^*$. Due to the rather small frequency of $F_\alpha$ that is hard to be analyzed in terms of the Dingle temperature and the Berry phase, we only did detailed analysis of $F_\beta$. For $F_\beta$, the Dingle temperature is $T_D$ = 83.3 K and the corresponding quantum scattering lifetime $\tau_Q$ = 1.461 × 10$^{-14}$ s, estimated from the inset of Fig. 2(e). Furthermore, the quantum mobility $\mu_Q$ = 524.179 cm$^2$ V$^{-1}$s$^{-1}$ obtained from $\mu_Q = e\tau_Q/m^*$. The results are summarized in Table I.

It is essential to achieve more insights into the topological state in LaAlSi. Dirac/Weyl system will produce a nontrivial $\varphi_B$ under a magnetic field, which could be probed by using the Landau level (LL) index fan diagram or a direct fit to the SdH oscillations by using the L-K formula. The phase shift is generally a sum as $\gamma - \delta$, as we mentioned above, where $\gamma$ is the phase factor expressed as 1/2 - $\varphi_B/2\pi$ and $\delta$ represents the dimension-dependent correction to the phase shift. In a two-dimensional (2D) case, $\delta$ amounts zero, while in a 3D case $\delta$ is ±1/8 where the sign depends on type of charge carriers and kind of cross-section extremum. Fig. 2(e) shows the plot of LL index $N$ as a function of 1/$B$ for the dominant frequency $F_\beta$. Here the $\Delta\rho_{xx}$ valley positions (closed circles) in 1/$B$ were assigned to be integer indices and the $\Delta\rho_{xx}$ peak positions (open circles) were assigned to be half-integer indices. All the points almost fall on a straight line, thus allowing a linear fit that gives an intercept of ~ 0.65, demonstrating a π Berry phase and suggesting the non-trivial state in LaAlSi.

Paying more attentions to the MR of LaAlSi, two more peaks at ~ 18 T and 32 T besides the two fundamental frequencies could also be identified and the stronger one at ~ 32 T can be clearly seen in the SdH oscillatory component, as marked by arrows shown by the insets to Figs. 2(a) and (b). To achieve more information, the MR vs. *B* is presented in Fig. 3(a) together with the form -d$^2\rho_{xx}$/d$B^2$, revealing hidden peaks marked by the dotted lines, which gradually disappear with increasing the temperature from 1.8 K to 30 K, indicative of a strong Zeeman effect which is also manifested by the broadening or even splitting of the oscillation peaks. As we mentioned above, in the L-K equation, the spin factor $R_s$ originates from lifting of the twofold spin degeneracy, so the L-K formula can be rewritten separately in the spin up and spin down cases [52]:

$$\Delta\rho \sim R_T R_D \left[ \cos\left[2\pi\left(\frac{F}{B}+\gamma-\delta+\frac{1}{2}\varphi\right) + 2\pi\left(\frac{F}{B}+\gamma-\delta-\frac{1}{2}\varphi\right)\right]\right]$$

where $\varphi = gm^*/2m_e$. According to the Landau fan diagram for both spin-up (red) and spin-down (black), a large Landé factor $g$ ~ 15.1 was obtained. Given such a large $g$ factor, it is reasonable to result in the Zeeman splitting effect as unveiled by the MR analysis. Moreover, the $g$ factor leads to the $R_S$ of ~ 0.4, indicating that there is no reversal appeared between the peak and valley positions of quantum oscillations, thus ensuring the accuracy of the above Berry phase analysis. However, three miscellaneous peaks marked by the red arrows in Fig. 3(a) are also visible, which might be caused by noise or new frequencies emerged under high magnetic field. Unfortunately, the numbers of miscellaneous peaks are rather few and the peaks are too weak to be analyzed further.

The results of first principles calculations for LaAlSi are summarized in Fig. 4. The obtained band structure without spin-orbit coupling (SOC), as presented in Fig. 4(a), shows that there are two bands crossing the $E_F$ and the conduction and valence bands cross each other along the Γ - Σ - Σ$_1$ path. The band structure indicates that LaAlSi has a semimetal ground state, which is also demonstrated by the calculated FS shown in Fig. 4(c)). It is noteworthy that the crossing between conduction and valence bands forms two pairs of "nodal rings" centered at $\Sigma$ and $\Sigma_1$ points on the $k_x = 0$ and $k_y = 0$ mirror planes, as schematically shown in Fig. 4(b).

When the SOC is considered, the nodal rings are gapped out as indicated in Fig. 4(d). Since LaAlSi has no inversion center, WPs are thus expected to exist in this compound. We find that there are three kinds of nonequivalent WPs, marked as $W_1$, $W_2$, and $W_3$, and the FSs arising from three Weyl fermion cones are marked in Fig. 4(c). The band structures around the three nonequivalent WPs are shown Figs. 4(e) - 4(g), from which we can see that $W_1$ and $W_3$ belong to type-I and locate ~ 0.1 eV above the $E_F$, meanwhile $W_2$ belongs to type-II and locates exactly on the $E_F$. The distributions of these three nonequivalent WPs in the momentum space are schematically shown in Figs. 4(i) and 4(j). $W_1$ and $W_3$ locate on the $k_z = 0$ plane and $W_2$ locates away from this plane. The coexistence of both type-I and type-II Weyl fermions and their positions in LaAlSi are very similar as the case of its sister compound LaAlGe, which was unveiled by ARPES measurements [30].

It is essential to probe the shape of the FS by magnetotransport measurements. The angle dependent MR is presented in Fig. 5(a) with the measurement geometry shown as the inset, which exhibits a clear angle dependent magnitude of the MR. Fig. 5(b) presents the angle dependent SdH oscillations with a constant offset after subtracting the smooth background, showing clear shift with the increase of $\theta$, i.e. the angle between $B$ and the $c$ axis. The $\theta$ dependence of frequencies derived from the SdH oscillations is shown in Fig. 5(c). The frequency of $F_\beta$ changes from 47.78 T at $\theta = 0°$ (out of plane) to 88.6 T or 125.71 T at $\theta = 70°$, unveiling the strong anisotropy of the Fermi pocket associated with the SdH oscillations. The splitting of $F_\beta$ at $70°$ may be caused by the sensitivity of the orbit to the magnetic field direction [33], which consequently produces the new frequency. As a contrast, the fundamental frequency $F_\alpha$ is almost angle independent. To achieve in-depth insights to the SdH oscillations, the angular dependence of the FS cross sectional areas for LaAlSi calculated by the first principles calculations are shown in Fig. 5(d). The remarkable agreement between the theoretical and experimental values suggests the reliability of the result. According to the calculations, $F_\alpha$ and $F_\beta$ are assigned to the electron pockets that are connected to the Weyl cones of $W_2$ and $W_1$, respectively, as marked in Fig. 4(c) and inset of Fig. 5(d), indicating that both type-I and type-II WSM states contribute to the magnetotransport properties of LaAlSi.

## IV. SUMMARY

In conclusion, our high magnetic field magnetotransport measurements and first principles calculations have unveiled both type-I and type-II Weyl states in the noncentrosymmetric LaAlSi, with all three types of nonequivalent Weyl points being very close to the Fermi level. This usual topological state naturally provides a unique platform for the study of multiple Weyl fermion physics as well as for the exploration of more intriguing topological phenomena. For example, There has theoretical predication that LaAlSi and LaAlGe may possess large intrinsic spin Hall effect and high conversion efficiency of charge-to-spin current by the spin-orbit torque [53]. Moreover, its sister compounds $R'AlPn$ ($R'$ = Ce and Pr, $Pn$ = Si, Ge) have been proposed as magnetic WSMs with both SI and TR symmetries being broken due to intrinsic magnetic order, which while still appeal experimental evidences. The exact role of interplay between the magnetism and electronic band structure remains unclear yet. In this regard, the nonmagnetic LaAlSi supports a comparison with the $R'AlPn$, which would favor the understanding about the role of magnetism in giving rise to the magnetic Weyl state.

**ACKNOWLEDEMENTS**

The authors acknowledge the support by the Major Research Plan of the National Natural Science Foundation of China (No. 92065201), the National Natural Science Foundation of China (Grant No. 11874264). Y.F.G. acknowledges the starting grant of ShanghaiTech University and the Program for Professor of Special Appointment (Shanghai Eastern Scholar) and the strategic Priority Research Program of Chinese Academy of Sciences (Grant No. XDA18000000). W.Z. is supported by the Shenzhen

Peacock Team Plan (Grant No. KQTD20170809110344233), and Bureau of Industry and Information Technology of Shenzhen through the Graphene Manufacturing Innovation Center (Grant No. 201901161514). The authors also thank the support from Analytical Instrumentation Center (#SPST-AIC10112914), SPST, ShanghaiTech University.

**References**

[1] N. P. Armitage, E. J. Mele, and A. Vishwanath, Rev. Mod. Phys. **90**, 015001 (2018).
[2] B. Yan and C. Felser, Annu. Rev. Cond. Mstter. Phys. **8**, 337 (2017).
[3] C. Zhang, H.-Z. Lu, S.-Q. Shen, Y. P. Chen, and F. Xiu, Sci. Bull. **63**, 580 (2018).
[4] X. Wan, A. M. Turner, A. Vishwanath, and S. Y. Savrasov, Phys. Rev. B **83**, 205101 (2011).
[5] H. Weng, C. Fang, Z. Fang, B. A. Bernevig, and X. Dai, Phys. Rev. X **5**, 011029(2015).
[6] S.-Y. Xu, *et al*. Science **349**, 613 (2015).
[7] A. A. Soluyanov, D. Gresch, Z. Wang, Q. Wu, M. Troyer, X. Dai, and B. A. Bernevig, Nature **527**, 495 (2015).
[8] K. Deng, *et al*. Nat. Phys. **12**, 1105 (2016).
[9] L. Huang, *et al*. Nat. Mater. **15**, 1155 (2016).
[10] A. Tamai, Q. S. Wu, I. Cucchi, F. Y. Bruno, S. Riccò, T. K. Kim, M. Hoesch, C. Barreteau, E. Giannini, C. Besnard, A. A. Soluyanov, and F. Baumberger, Phys. Rev. X **6**, 031021 (2016).
[11] J. Jiang, Z. Liu, Y. Sun, H. Yang, C. Rajamathi, Y. Qi, L. Yang, C. Chen, H. Peng, et al. Nat. Commun. **8**, 13973(2017).
[12] M.-Y. Yao, N. Xu, Q. S. Wu, G. Autès, N. Kumar, V. N. Strocov, N. C. Plumb, M. Radovic, O. V. Yazyev, C. Felser, J. Mesot, and M. Shi, Phys. Rev. Lett. **122**, 176402(2019).
[13] C. Shekhar, A. K. Nayak, Y. Sun, M. Schmidt, M. Nicklas, I. Leermakers, U. Zeitler, Y. Skourski, J. Wosnitza, and Z. Liu, Nat. Phys. **11**, 645 (2015).
[14] M. N. Ali, J. Xiong, S. Flynn, J. Tao, Q. D. Gibson, L. M. Schoop, T. Liang, N. Haldolaarachchige, M. Hirschberger, and N. P. Ong, Nature **514**, 205 (2014).
[15] Z. Zhu, X. Lin, J. Liu, B. Fauqué, Q. Tao, C. Yang, Y. Shi, and K. Behnia, Phys. Rev. Lett. **114**, 176601 (2015).
[16] C.-L. Zhang, S.-Y. Xu, I. Belopolski, Z. Yuan, Z. Lin, B. Tong, G. Bian, N. Alidoust, C.-C. Lee, and S.-M. Huang, Nat. Commun. **7**, 1 (2016).
[17] X. Huang, L. Zhao, Y. Long, P. Wang, D. Chen, Z. Yang, H. Liang, M. Xue, H. Weng, Z. Fang, X. Dai, and G. Chen, Phys. Rev. X **5**, 031023 (2015).
[18] J. Xiong, S. K. Kushwaha, T. Liang, J. W. Krizan, M. Hirschberger, W. Wang, R. J. Cava, and N. P. Ong, Science **350**, 413 (2015).
[19] K.-Y. Yang, Y.-M. Lu, and Y. Ran, Phys. Rev. B **84**, 075129 (2011).

[20] E. Liu, Y. Sun, N. Kumar, L. Muechler, A. Sun, L. Jiao, S.-Y. Yang, D. Liu, A. Liang, and Q. Xu, Nat. Phys. **14**, 1125 (2018).
[21] Q. Wang, Y. Xu, R. Lou, Z. Liu, M. Li, Y. Huang, D. Shen, H. Weng, S. Wang, and H. Lei, Nat. Commun. **9**, 1 (2018).
[22] S. Zhong, J. Orenstein, and J. E. Moore, Phys. Rev. Lett. **115**, 117403 (2015).
[23] D. F. Liu, *et al*. Science **365**, 1282 (2019).
[24] Y. Araki, and K. Nomura, Phys. Rev. Appl. **10**, 014007 (2018).
[25] H. Su, *et al*. APL Mater. **8**, 011109 (2020).
[26] J.-R. Soh, F. de Juan, M. G. Vergniory, N. B. M. Schroter, M. C. Rahn, D. Y. Yan, J. Jiang, M. Bristow, P. Reiss, J. N. Blandy, Y. F. Guo, Y. G. Shi, T. K. Kim, A. McCollam, S. H. Simon, Y. Chen, A. I. Coldea, A. T. Boothroyd. Phys. Rev. B **100**, 201102(R) (2019).
[27] M. Yao, H. Lee, N. Xu, Y. Wang, J. Ma, O. V. Yazyev, Y. Xiong, M. Shi, G. Aeppli, Y. Soh, arXiv:1810.08514.
[28] I. Belopolski, *et al*. Science **365**, 1278 (2019).
[29] N. Morali, *et al*. Science **365**, 1286 (2019).
[30] S.-Y. Xu, N. Alidoust, G. Chang, H. Lu, B. Singh, I. Belopolski, D. S. Sanchez, X. Zhang, G. Bian, and H. Zheng, Sci. Adv. **3**, e1603266 (2017).
[31] D. S. Sanchez, G. Chang, I. Belopolski, H. Lu, J.-X. Yin, N. Alidoust, X. Xu, T. A. Cochran, X. Zhang, Y. Bian, S. S. Zhang, Y.-Y. Liu, J. Ma, G. Bian, H. Lin, S.-Y. Xu, S. Jia, and M. Z. Hasan, Nat. Commun. **11**, 3356 (2020).
[32] K. Zhang, T. Wang, X. Pang, F. Han, S.-L. Shang, N. T. Hung, Z.-K. Liu, M. Li, R. Saito, and S. Huang, Phys. Rev. B **102**, 235162 (2020).
[33] H.-Y. Yang, B. Singh, J. Gaudet, B. Lu, C.-Y. Huang, W.-C. Chiu, S.-M. Huang, B. Wang, F. Bahrami, and B. Xu, arXiv: 2006.07943 (2020).
[34] H. Hodovanets, C. J. Eckberg, P. Y. Zavalij, H. Kim, W.-C. Lin, M. Zic, D. J. Campbell, J. S. Higgins, and J. Paglione, Phys. Rev. B **98**, 245132 (2018).
[35] P. Puphal, V. Pomjakushin, N. Kanazawa, V. Ukleev, D. J. Gawryluk, J. Z. Ma, M. Naamneh, N. C. Plumb, L. Keller, R. Cubitt, E. Pomjakushina, J. S. White, Phys. Rev. Lett. **124**, 017202 (2020).
[36] J. Gaudet, H.-Y. Yang, S. Baidya, B. Z. Lu, G. Y. Xu, Y. Zhao, J. A. Rodriguez, C. M. Hoffmann, D. E. Graf, D. H. Torchinsky, P. Nikoli, D. Vanderbilt, F. Tafti, C.L. Broholm, arXiv:2012.12970 (2020).
[37] P. Puphal, C. Mielke, N. Kumar, Y. Soh, T. Shang, M. Medarde, J. S. White, and E. Pomjakushina, Phys. Rev. Mater. **3**, 024204 (2019).
[38] G. Chang, B. Singh, S.-Y. Xu, G. Bian, S.-M. Huang, C.-H. Hsu, I. Belopolski, N. Alidoust, D. S. Sanchez, H. Zheng, H. Lu, X. Zhang, Y. Bian, T.-R. Chang, H.-T. Jeng, A. Bansil, H. Hsu, S. Jia, T. Neupert, H. Lin, and M. Z. Hasan, Phys. Rev. B **97**, 041104(R) (2018).
[39] P. E. Blöchl, Phys. Rev. B **50**, 17953 (1994).
[40] J. Lehtomäki, I. Makkonen, M. A. Caro, and O. Lopez-Acevedo, J. Chem. Phys. **141**, 234102 (2014).
[41] J. P. Perdew, K. Burke, and M. Ernzerhof, Phys. Rev. Lett. **77**, 3865 (1996).
[42] J. P. Perdew and Y. Wang, Phys. Rev. B **46**, 12947 (1992).

[43] G. Kresse and J. Hafner, Phys. Rev. B **47**, 558 (1993).

[44] G. Kresse and J. Furthmüller, Comput. Mater. Sci. **6**, 15 (1996).

[45] G. Kresse and J. Furthmüller, Phys. Rev. B **54**, 11169 (1996).

[46] A. A. Mostofi, J. R. Yates, Y.-S. Lee, I. Souza, D. Vanderbilt, and N. Marzari, Comput. Phys. Commun. **178**, 685 (2008).

[47] N. Marzari and D. Vanderbilt, Phys. Rev. B **56**, 12847 (1997).

[48] I. Souza, N. Marzari, and D. Vanderbilt, Phys. Rev. B **65**, 035109 (2001).

[49] Q. Wu, S. Zhang, H.-F. Song, M. Troyer, and A. A. Soluyanov, Comput. Phys. Commun. **224**, 405 (2018).

[50] M. L. Sancho, J. L. Sancho, J. L. Sancho, and J. Rubio, J. Phys. F: Met. Phys. **15**, 851 (1985).

[51] E. Lifshits and A. Kosevich, J. Phys. Chem. Solids **4**, 1 (1958).

[52] Y. Liu, X. Yuan, C. Zhang, Z. Jin, A. Narayan, C. Luo, Z. Chen, L. Yang, J. Zou, and X. Wu, Nat. Commun. **7**, 1 (2016).

[53] T. N. Yu, J. Yuan, Y. Luo, Y. Wu, and L. Shen, arXiv:2005.05560 (2020).

**TABLE I.** Parameters derived from SdH oscillations for LaAlSi.

| $F$ (T) | $S$ (nm$^{-2}$) | $K_F$ (nm$^{-1}$) | $V_F$ (ms$^{-1}$) | $E_F$ (meV) | $m^*/m_e$ | Berry phase | $T_D$ (K) | $\mu$ (cm$^2$/Vs) |
|---|---|---|---|---|---|---|---|---|
| 7.96 | 0.0758 | 0.1554 | 400337 | 41.019 | 0.045 | - | - | - |
| 47.78 | 0.455 | 0.381 | 900758 | 226.117 | 0.049 | 1.3 π | 83.3 | 524.179 |

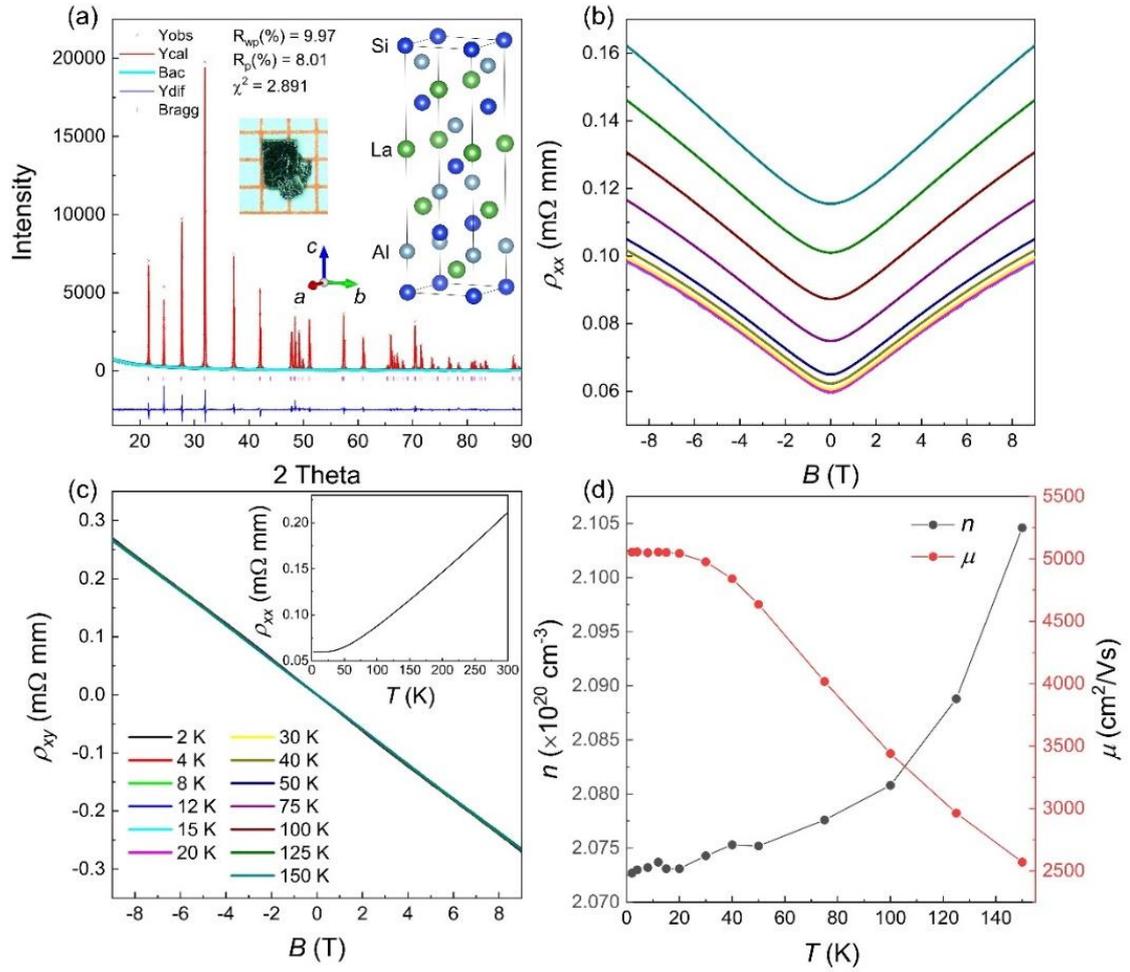

**Fig. 1.** (a) Powder x-ray diffraction and Rietveld refinement result of LaAlSi. Insets: an image of a typical single crystal with clean (001) plane and the schematic crystal structure. (b) and (c) Longitudinal $\rho_{xx}$ and Hall resistivity versus low magnetic field $B$ in the temperature range of 2 – 150 K. Inset of (c) shows the longitudinal resistivity versus temperature. (d) The calculated carrier density and mobility versus temperature.

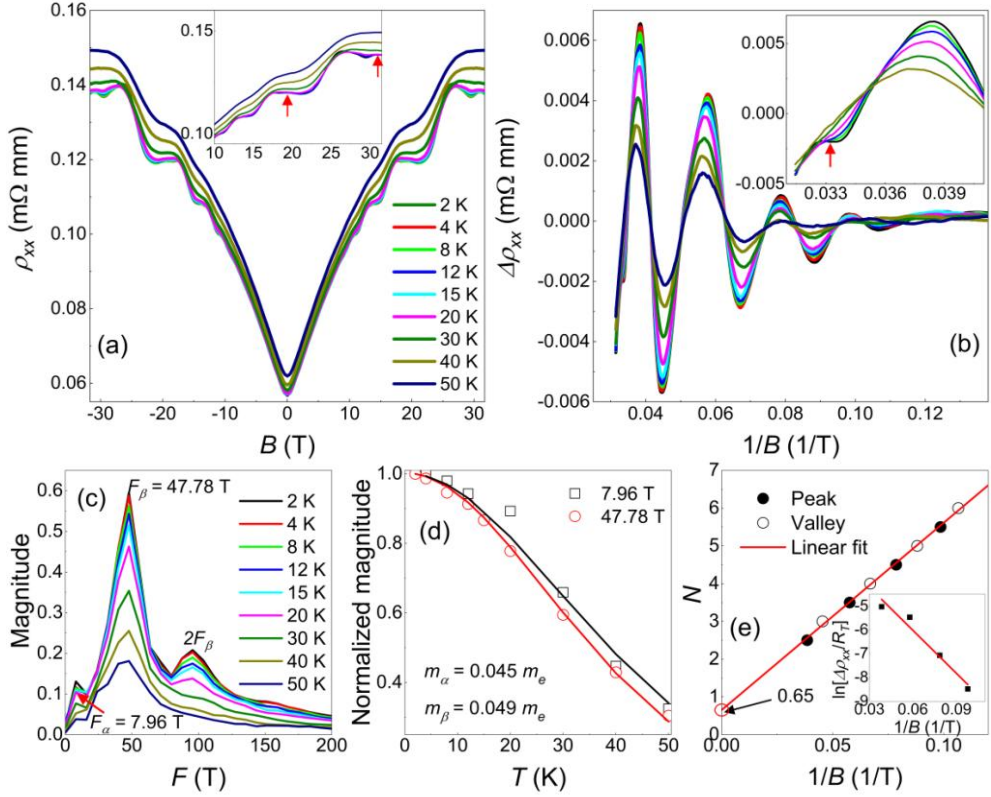

**Fig. 2.** (a) The longitudinal magnetoresistance vs. magnetic field, which displays clear quantum oscillations under high magnetic field. Inset shows two more peaks at ~ 18 T and 32 T in addition to the main peaks. (b) The SdH oscillatory component as a function of $1/B$ after subtracting the background. Inset shows one more peak at ~ 32 T in addition to the main peaks. (c) *FFT* spectra of $\Delta\rho_{xx}$ shows the obtained two fundamental frequencies $F_\alpha$ = 7.96 T and $F_\beta$ = 47.78 T. (d) Temperature dependence of relative *FFT* magnitude of the two fundamental frequencies. The solid lines denote the fitting by using the L-K formula. (e) Landau level indices of $F_\beta$ extracted from the SdH oscillations plotted as a function of $1/B$. The solid line indicates the linear plots to the data. Inset: Dingle plot of the SdH oscillation at $T$ = 2 K.

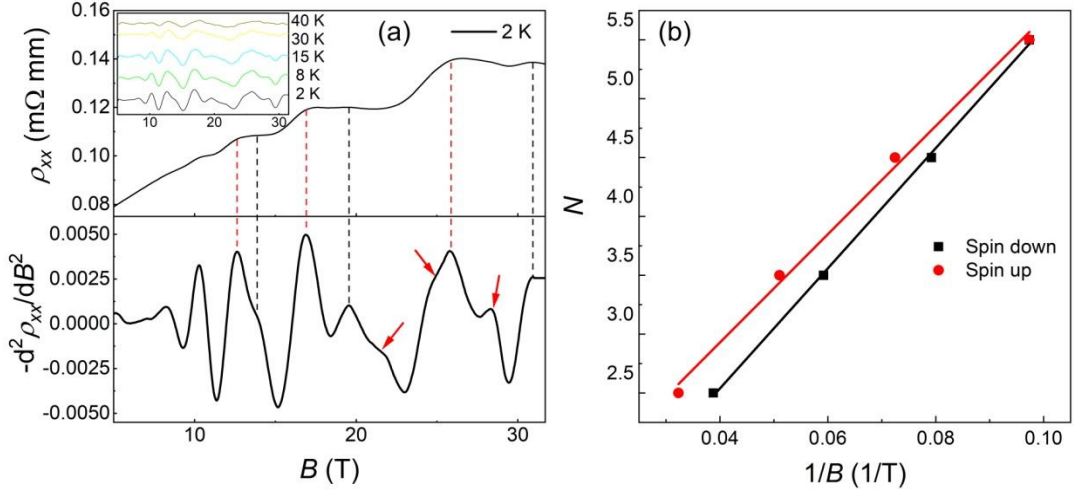

**Fig. 3.** (a) Longitudinal $\rho_{xx}$ and $-d^2\rho_{xx}/dB^2$ oscillation component versus magnetic field $B$ at 2 K. Inset: $-d^2\rho_{xx}/dB^2$ oscillation component at different $T$. (b) Landau fan diagram for the $F_\beta$ with Zeeman splitting.

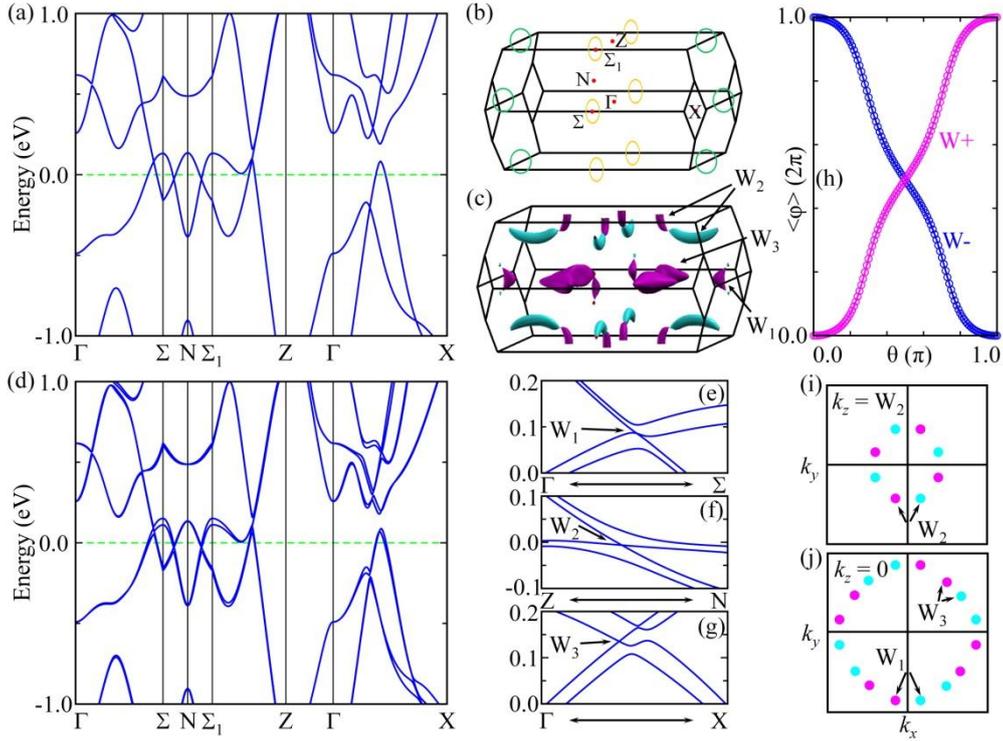

**Fig. 4.** (a) Electronic band structure of LaAlSi without including the SOC. (b) Schematic illustration of the nodal rings in the Brillouin zone. (c) Fermi surfaces of LaAlSi over the bulk Brillouin zone. (d) Electronic band structure of LaAlSi with including the SOC. (e)-(g) Band structures near the Weyl points $W_1$, $W_2$, and $W_3$. (h) Evolution of the Wannier charge centers

around the sphere enclosing W+ and W-. (i) and (j) Weyl points in LaAlSi at $k_z = 0$ and $k_z = W_2$ planes labeled by cyan/magenta dots with different chirality.

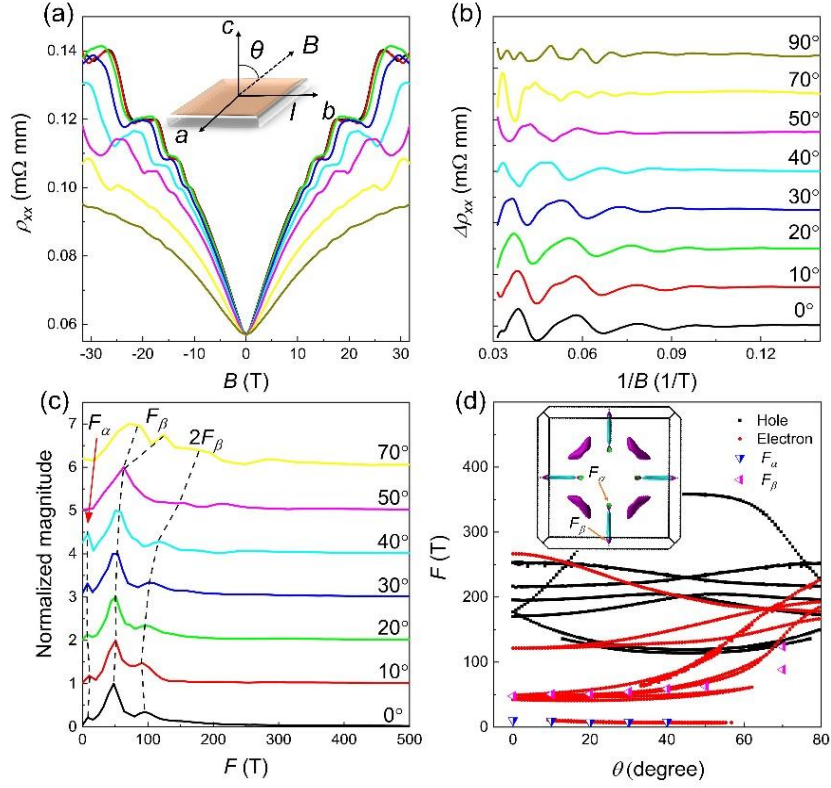

**Fig. 5.** (a) Longitudinal $\rho_{xx}$ versus magnetic field $B$ at different angles between $B$ and the $c$ axis at 2 K. Inset shows the schematic measurement configuration. (b) SdH oscillatory component as a function of $1/B$ at different angles. (c) FFT spectra of $\Delta\rho_{xx}$ at different angles. (d) Quantum oscillation frequencies as a function of the angle $\theta$ for the hole and electron FSs derived from experiments and calculations. Inset: Projection of three dimensional FSs on the $k_x$-$k_y$ plane. The $F_\alpha$ and $F_\beta$ are marked by red arrows, which are connected to the Weyl cones corresponding to the $W_2$ (green) and $W_1$ (purple) Weyl points, respectively.